\documentstyle[11pt,newpasp,twoside,epsf]{article}
\def\edcomment#1{\iffalse\marginpar{\raggedright\sl#1\/}\else\relax\fi}
\reversemarginpar

\begin{document}

\title{
Extrasolar Terrestrial Planets: Can We Detect Them Already?}

\author{Michael Endl}
\affil{
McDonald Observatory, The University of Texas at Austin, Austin, TX 78712, USA}

\author{Martin K\"urster}
\affil{
Th\"uringer Landessternwarte Tautenburg, Sternwarte 5,
              07778 Tautenburg, Germany}

\author{Fr\'ed\'eric Rouesnel}
\affil{
UFR DESS - Observatory of Paris-Meudon,
F-92195 Meudon cedex, France
}

\author{Sebastian Els}
\affil{
Isaac Newton Group of Telescopes, Apartado de Correos 321,
                E-38700 Santa Cruz de La Palma, Spain}

\author{Artie P. Hatzes}
\affil{
Th\"uringer Landessternwarte Tautenburg, Sternwarte 5,
              07778 Tautenburg, Germany}

\author{William D. Cochran}
\affil{
McDonald Observatory, The University of Texas at Austin, Austin, TX 78712, USA}

\begin{abstract}
With the example of Proxima Centauri we discuss the
feasibility of detecting terrestrial planets (1 to a few ${\rm M}_{\oplus}$) 
using the high precision radial velocity ($RV$) technique. If a
very high $RV$ precision for M stars is achieved even planets
with these extremely low masses become detectable. For Proxima
Cen (M5V), one of the prime targets of our M-stars planet search program using the
UVES spectrograph \& iodine cell at the ESO VLT UT2, we obtain a
long term $RV$ precision of $2.5~{\rm m\,s}^{-1}$. Based on numerical
simulations we determine that this level of precision would have 
already allowed us to detect planets with $m\sin i= 4$ to $6~{\rm M}_{\oplus}$ inside
the habitable zone of Proxima Cen.    
\end{abstract}

\section{Introduction}

All extrasolar planets orbiting main-sequence stars known to date are giant planets ranging in
mass from $m \sin i = 0.12~{\rm M}_{\rm Jupiter}$ to $17~{\rm M}_{\rm Jup}$ (with $i$ the unknown
angle between the orbital plane and the sky).
It is virtually impossible to detect terrestrial planets (rocky objects of 1 to a few ${\rm M}_{\oplus}$)
around F, G and K-type stars using the high-precision $RV$ technique, since Earth-mass planets induce
only negligible reflex-motions on their host stars.
This scenario changes in the faint and low-mass regime of the
Hertzsprung-Russell diagram. In the case of M-dwarf stars the low stellar primary mass leads to detectable
$RV$ amplitudes even for planets of a few ${\rm M}_{\oplus}$ and less in short-period orbits.
Fig.1 shows the $RV$ signatures of planets with $m \sin i = 1$ to $2.5~{\rm M}_{\oplus}$ orbiting an M5V star with
a mass of $0.1~{\rm M}_{\sun}$. Due to its intrinsic faintness the habitable zone is located 
very close to the star (Kasting, Whitmire, \& Reynolds 1993). This, too, favors the detection by the $RV$ technique, 
since shorter periods also mean higher $RV$ amplitudes. Although these nearby planets are probably
tidally locked into synchronous rotation, a 3-D climate study by Joshi, Haberle, \& Reynolds (1997) demonstrated
that these planets are still likely to be habitable.      

\begin{figure}
\plotfiddle{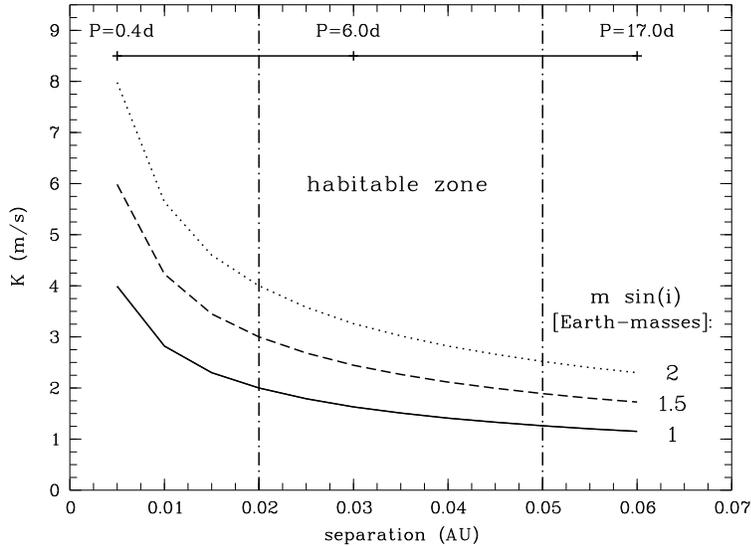}{7.0cm}{270}{40}{40}{-180}{220}
\caption{Simulation of radial velocity variations ($RV$ semi-amplitude K) for a low-mass
M-dwarf ($0.1~{\rm M}_{\sun}$) due to orbiting terrestrial planets plotted vs. 
orbtial separation. The three curves display the K-amplitudes for planets with 
$m \sin i = 1~{\rm M}_{\oplus}$ (solid), $1.5~{\rm M}_{\oplus}$ (dashed) and $2~{\rm M}_{\oplus}$
(dotted curve) residing in circular orbits. 
The vertical dashed lines show the borders of the habitable zone for this type of star after
Kasting et al. (1993). 
}
\end{figure}

\section{Searching for terrestrial planets inside the habitable zone of \\Proxima Centauri}

For the past two years we have used the ESO VLT and the UVES spectrograph to monitor a sample
of M-dwarf stars in the southern hemisphere to search for planetary companions. Our 
$RV$ results for Proxima Cen (M5V, $V=11.05$), obtained
with an enhanced version of our {\it Austral} radial velocity code (Endl, K\"urster, \& Els 2000), 
are displayed in Fig.2. 
Over the time span of more than one year the
star experiences a linear acceleration of unknown origin. 
The rms-scatter around this trend is $2.5~{\rm m\,s}^{-1}$. 

\begin{figure}
\plotfiddle{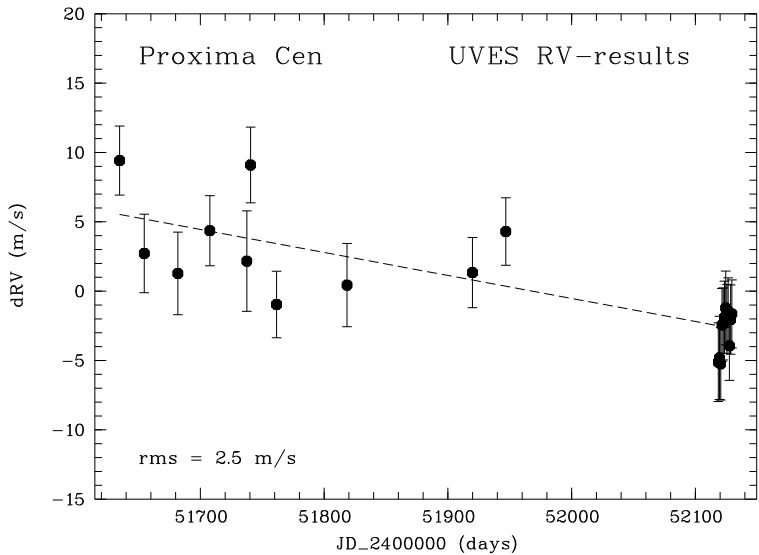}{7.0cm}{270}{40}{40}{-180}{220}
\caption{Differential radial velocity measurements of Proxima Cen obtained with the
UVES spectrograph \& iodine cell at the ESO VLT UT2 telescope. We find a linear
acceleration of unknown origin (dashed line), the rms-scatter around this trend
is $2.5~{\rm m\,s}^{-1}$. 
}
\end{figure}   

\subsection{Detection sensitivity}

We then take the 22 RV-measurements (after subtraction of the linear trend) and 
determine our planet detection threshold in the period range for orbits inside the habitable
zone (2 to 16 days), using the technique from Endl at al. (2001). 
For the primary mass we adopt a value of $0.11~{\rm M}_{\sun}$ after Henry et al. (1999).
Fig.3 shows our momentary detection capability based on the
$RV$ data collected so far. With a confidence of $> 99\%$ we could have already detected {\it all}
planets with $m\sin i$ values higher than $4$ to $6~{\rm M}_{\oplus}$ inside the habitable zone. 
With a chance probability of $\approx 50\%$ (i.e. only half of the test signals were recovered 
successfully with a confidence of $> 99\%$) we would have even found signals of planets     
down to $m\sin i \approx 2.5~{\rm M}_{\oplus}$. 
These new constraints on planets inside the habitable zone of Proxima Cen 
supplement the existing limits for giant planets around this star by our
ESO CES planet search program (K\"urster et al. 1999) and the HST Fine 
Guidance Sensor astrometric results of Benedict et al. (1999). 

\begin{figure}
\plotfiddle{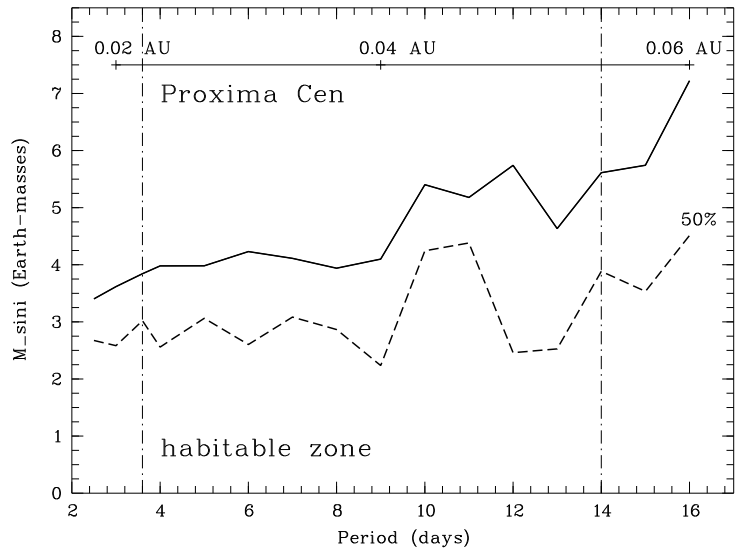}{7.0cm}{270}{40}{40}{-180}{220}
\caption{Detection threshold for low-mass planets orbiting inside the habitable zone of
Proxima Cen. The solid thick line denotes the limit where we could have detected {\it all}
planets with a confidence of $> 99\%$. Hence we can exclude the presence of planets with $m\sin i$ values
on and above this line. The dashed line shows the limits where we had a $50\%$ chance of detection.
}
\end{figure}

\section{Conclusions}
Measuring radial velocities of M-dwarfs with the appropriate high level of precision
allows us to detect extremely low-mass planets in short period orbits. With the 
22 $RV$ measurments we obtained so far for Proxima Cen, we could have already detected
{\it all} planets with $m\sin i= 4$ to $6~{\rm M}_{\oplus}$ inside the habitable zone.
So the answer to the title question is: {\it not yet}, but we get pretty close, and 
our sensitivity will further improve by extending the monitoring time span.

\acknowledgements
This work is supported by NASA Grant NAG5-9227 and by NSF Grant AST-9808980. We
thank the ESO OPC and DDTC for generous allocation of observing time to our VLT M-dwarf
planet search program (Proposals 69.C-0722, 68.C-0415, 67.C-5700).

\end{document}